\documentclass[seceq]{ptptex}
\usepackage[dvips]{graphicx}

\preprintnumber[3cm]{
KUNS-2320\\PTPTeX ver.0.8\\ August, 1997}

\title{Fatal Effects of Charges on Stability of Black Holes\\ in Lovelock Theory
}

\author{Tomohiro \textsc{Takahashi} }

\inst{
Department of Physics,  Kyoto University, Kyoto, 606-8502, Japan
}

\abst{%
We study the stability of static spherically symmetric charged black holes under tensor type perturbations 
in Lovelock theory which is a natural higher dimensional generalization of Einstein theory.  
We derive the master equation for tensor type perturbations and present the criteria for the stability. 
Examining these criteria numerically, 
we show that a black hole with small charge has the instability if its mass is as small as extreme mass. 
Combined with our previous result that neutral black hole has no dynamical instability in odd dimensions, 
this result suggests that charges have fatal effects on black holes in Lovelock theory. 
}

\begin{document}

\maketitle

\section{Introduction}

One of the most exciting predictions of the braneworld with large extra-dimensions 
is the possibility of higher dimensional black hole creation at  the LHC. \cite{Giddings:2001bu} \  
 Then, it is important to study higher dimensional black holes. 
 Especially, stability analysis of higher dimensional black holes is important 
 because unstable black hole solutions are not attractor solutions; that is, 
 black holes with instability must not be created at the LHC.   

The stability of higher dimensional black holes in Einstein theory has been 
intensively studied. The most famous one is stability  analysis 
of higher dimensional Schwarzschild black hole by Kodama and Ishibashi.\cite{Kodama:2003jz}\ 
They have shown that  higher dimensional Schwarzschild black holes are stable for all type perturbations. 
They have also extended their works to charged black holes in higher dimensions and derived master equations.\cite{Kodama}\ 
Furthermore, for charged black holes, numerical studies have also been done.\cite{Konoplya:2007jv}\ 
For the stability analysis of rotating black holes in higher dimensions,
a group theoretical method is developed,\cite{Murata:2007gv} \ 
which is used for 5-dimensional rotating black holes with equal angular momenta\cite{Murata:2008yx} \ 
and also used to study the stability of squashed black holes.\cite{Kimura:2007cr,Ishihara:2008re,Nishikawa:2010zg}\ 
The stability of a special class of rotating black holes in more than 5-dimensions is also
studied. \cite{Kunduri:2006qa,Oota:2008uj,Kodama:2008rq} \ 

As mentioned above, many studies have been done based on Einstein theory. 
It is no less important  than these 
to examine the stability of higher dimensional black holes in more general gravitational theories. 
This is because Einstein theory is not the most general gravitational theory
 which contains terms only up to the second order derivatives of metric 
in the equations of motion. 
The most general theory in this sense is Lovelock theory.\cite{Lovelock:1971yv} \  
Furthermore, 
because black hole productions occur at the fundamental scale, 
 Einstein theory is not reliable any more. 
 In fact, it is known Einstein theory is only a low energy limit of string theory.\cite{Boulware:1985wk} \ 
 In string theory, there are higher curvature corrections in addition to the
Einstein-Hilbert term.\cite{Boulware:1985wk} \ 
Thus, it is natural to extend gravitational theory into those
 with higher curvature corrections in higher dimensions.
 Lovelock theory  belongs to such class of theories.\cite{Lovelock:1971yv,Christos} \ 
Then, it is worthy to  extend the stability analysis to this general Lovelock theory.

In Lovelock theory, it is known that
there exist static spherical symmetric black hole solutions~\cite{Christos,Wheeler:1985nh} 
(and also topological black hole solutions are found~\cite{Cai:2001dz}). 
For this static spherical symmetric black holes, stability analyses under all type 
perturbations have been performed.~\cite{Dotti:2004sh,Neupane:2003vz,Gleiser:2005ra,Takahashi:2009dz}
It is shown that there exists the scalar mode instability in odd dimensions,   
the tensor mode instability in even dimensions and no instability under vector type perturbation in all dimensions. 
In 2nd order Lovelock theory, stability analysis is extended to black string.~\cite{CuadrosMelgar:2010cc}\ 
They have shown that scalar perturbations have an  
exponentially decaying behavior under s-mode approximations. 

However, because black holes may be produced by proton-proton collision at the LHC, 
it is natural to suppose these black holes have charges. In Lovelock-Maxwell theory, 
there exist static spherical symmetric charged black hole solution, namely charged Lovelock black hole solution. \cite{Christos}\   
Then, it is important to extend the above discussion to this charged solution. 

In this paper, 
we study the linear stability of charged Lovelock black hole solutions. 
Our purpose is examining the response of instability  to the charge and 
considering the black holes creation in Lovelock theory at LHC. 
In this paper, for the first step, we only concentrate on tensor type perturbation . 
Furthermore, we mainly consider  the second order and third order Lovelock theory for simplicity, 
although  it is important to consider higher Lovelock terms. \cite{Rychkov:2004sf} \ 

The organization of this paper is as follows.
 In section \ref{seq:2}, we review Lovelock theory and confirm the existence of asymptotically flat black hole solutions. 
In section \ref{seq:3}, we consider tensor perturbations and denote the condition for stability of charged Lovelock black holes. 
The method using in this section is the same as our previous paper.\cite{Takahashi:2009dz} \  
In section \ref{seq:4}, we check the criteria presented in section \ref{seq:3} numerically. 
We mainly examine the second order and third order Lovelock theory in this section. 
 The final section \ref{seq:5} is devoted to the conclusion. 
 
\section{Charged Lovelock Black Hole Solutions}
\label{seq:2}
In this section, we review Lovelock theory and introduce 
charged black hole solutions. 

In Ref.~\citen{Lovelock:1971yv}, the most general symmetric, divergence free rank (1,1) tensor 
is constructed out of a metric and its first and second derivatives.
The corresponding Lagrangian can be constructed from $m$-th order Lovelock terms
\begin{eqnarray}
  {\cal L}_m = \frac{1}{2^m} 
  \delta^{\lambda_1 \sigma_1 \cdots \lambda_m \sigma_m}_{\rho_1 \kappa_1 \cdots \rho_m \kappa_m}
  R_{\lambda_1 \sigma_1}{}^{\rho_1 \kappa_1} \cdots  R_{\lambda_m \sigma_m}{}^{\rho_m \kappa_m}
                       \ ,
\end{eqnarray}
where  $R_{\lambda \sigma}{}^{\rho \kappa}$ is the Riemann tensor in $D$-dimensions
and $\delta^{\lambda_1 \sigma_1 \cdots \lambda_m \sigma_m}_{\rho_1 \kappa_1 \cdots \rho_m \kappa_m}$ is the 
generalized totally antisymmetric Kronecker delta defined by  
\begin{eqnarray}
\delta^{\mu_1\mu_2\cdots \mu_p}_{\nu_1\nu_2\cdots\nu_p}={\rm det}
\left(
\begin{array}{cccc}
\delta^{\mu_1}_{\nu_1}&\delta^{\mu_1}_{\nu_2}&\cdots&\delta^{\mu_1}_{\nu_p}\\
\delta^{\mu_2}_{\nu_1}&\delta^{\mu_2}_{\nu_2}&\cdots&\delta^{\mu_2}_{\nu_p}\\
\vdots&\vdots&\ddots&\vdots\\
\delta^{\mu_p}_{\nu_1}&\delta^{\mu_p}_{\nu_2}&\cdots&\delta^{\mu_p}_{\nu_p}
\end{array}
\right)\ .
\nonumber
\end{eqnarray}
Then, Lovelock Lagrangian in  $D$-dimensions is defined by
\begin{eqnarray}
  L = \sum_{m=0}^{k} b_m {\cal L}_m \ ,   \nonumber
\end{eqnarray}
where we defined the maximum order $k\equiv [(D-1)/2]$ and  $b_m$ are 
arbitrary constants. 
Here, $[z]$ represents the maximum integer satisfying $[z]\leq z$. 
Then, fixing $k$, $n=D-2$ satisfies $n=2k-1$ or $n=2k$.
Hereafter, we set $b_0=-2\Lambda$, $b_1=a_1=1$ and $b_m=a_m/m\ (m\geq 2)$, for convenience.  

Then, the action for Lovelock-Maxwell system is 
\begin{eqnarray}
S=\int \sqrt{-g}\left[-2\Lambda+\sum_{m=1}^k\left\{\frac{a_m}{m}{\cal L}_m\right\}\right]-\int \sqrt{-g}\frac{1}{4}F_{\mu\nu}F^{\mu\nu},
\label{total_action}
\end{eqnarray}
where $F_{\mu\nu}=A_{\mu;\nu}-A_{\nu;\mu}$ means field strength of electromagnetic field and $A_{\mu}$ is vector potential.  
Taking variation of this action with respect to $g_{\mu\nu}$, we can derive   
\begin{eqnarray}
{\cal G}_{\mu}{}^{\nu}=T_{\mu}{}^{\nu} ,
\label{}
\end{eqnarray}
where ${\cal G}_{\mu}{}^{\nu}$ is Lovelock tensor defined as 
\begin{eqnarray}
{\cal G}_{\mu}{}^{\nu}=\Lambda \delta_{\mu}^{\nu}-\sum_{m=1}^{k}\frac{1}{2^{m+1}}\frac{a_m}{m} 
	 \delta^{\nu \lambda_1 \sigma_1 \cdots \lambda_m \sigma_m}_{\mu \rho_1 \kappa_1 \cdots \rho_m \kappa_m}
       R_{\lambda_1 \sigma_1}{}^{\rho_1 \kappa_1} \cdots  R_{\lambda_m \sigma_m}{}^{\rho_m \kappa_m}
\label{EOM}
\end{eqnarray}
and 
\begin{eqnarray}
T_{\mu}{}^{\nu}=F_{\mu\lambda}F^{\nu\lambda}-\frac{1}{4}F_{\lambda\rho}F^{\lambda\rho}\delta_{\mu}^{\nu}
\label{}
\end{eqnarray}
is energy momentum tensor of Maxwell field. 
Furthermore, by varying (\ref{total_action}) by $A_{\mu}$, we can get parts of  Maxwell equations 
\begin{eqnarray}
F^{\mu\nu}{}_{;\nu}=0, 
\label{maxwell}
\end{eqnarray}
and the rests of Maxwell equations can be derived from the identity $dF=0$, which means 
\begin{eqnarray}
F_{[\mu\nu;\lambda]}=0.
\label{maxwell_bianchi}
\end{eqnarray}

As shown in Ref \citen{Christos}, 
there exist static spherical symmetric solution of these equations. 
Let us consider the following metric 
\begin{eqnarray}
ds^2=-f(r)dt^2+1/f(r)dr^2+r^2\gamma_{ij}dx^idx^j,
\label{metric_ansatz}
\end{eqnarray}
where $\gamma_{ij}$ means metric of $S^n$ and $n=D-2$. 
Using this, Lovelock tensor is calculated as follows:
\begin{eqnarray}
	&\ &{\cal G}_t^t={\cal G}_r^r=-\frac{n}{2r^n}\left(r^{n+1}W[\psi]\right)^{\prime}\ ,\nonumber\\
	             &\ &{\cal G}_i^j=-\frac{1}{2r^{n-1}}\left(r^{n+1}W[\psi]\right)^{\prime\prime}\ ,\nonumber\\
	             &\ &{\rm otherwise}=0\ , \label{1}
\end{eqnarray}
where $\psi$ is defined as 
\begin{eqnarray}
	f(r)=1-r^2 \psi(r) \nonumber\label{}
\end{eqnarray}
and $W[\psi]$ is 
\begin{eqnarray}
\left\{
\begin{array}{l}
	W[\psi]\equiv\sum_{m=2}^{k}\left[\frac{\alpha_m}{m}\psi^m\right]+\psi-\frac{2\Lambda}{n(n+1)} \\
	\alpha_m=a_m\left\{\prod_{p=1}^{2m-2}(n-p)\right\}
	\end{array}
	\right .
	\ .\label{}
\end{eqnarray}

For Maxwell field, we assume spherical symmetric electric field, that is,  
\begin{eqnarray}
F^{tr}=E(r),\quad {\rm otherwise}=0\ .
\label{max_ansatz}
\end{eqnarray}
Substituting this ansatz for (\ref{maxwell}) shows that $E(r)$ satisfies 
\begin{eqnarray}
\partial_r(r^nE(r))=0\Rightarrow E(r)=\sqrt{n(n-1)}Q/r^n\ ,
\label{max_sol}
\end{eqnarray}
where $\sqrt{n(n-1)}Q$ is constant of integral which means charge of black hole. 
The normalization factor $\sqrt{n(n-1)}$ is only for convenience. 
Note that (\ref{max_ansatz}) and (\ref{max_sol}) satisfies the identity (\ref{maxwell_bianchi}). 
Hence, this is exact solution for Maxwell equations.  The energy-momentum tensor for this solution is 
\begin{eqnarray}
T_t^t=T_r^r=-\frac{E^2}{2}&=&-\frac{n(n-1)Q^2}{2r^{2n}}\ ,\nonumber\\
	      T_i^j=\frac{E^2}{2}\delta_i^j&=&\frac{n(n-1)Q^2}{2r^{2n}}\delta_i^j\ , \nonumber\\
	      {\rm otherwise}&=&0\ .\label{2}
\label{}
\end{eqnarray}

The results (\ref{1}) and (\ref{2}) leads Lovelock equation (\ref{EOM})  as  
\begin{eqnarray}
 -\frac{n}{2r^n}\left(r^{n+1}W[\psi]\right)^{\prime}&=&-\frac{1}{2}\frac{n(n-1)Q^2}{r^{2n}}\ ,\nonumber\\
	-\frac{1}{2r^{n-1}}\left(r^{n+1}W[\psi]\right)^{\prime\prime}&=&\frac{1}{2}\frac{n(n-1)Q^2}{r^{2n}}\ ,\nonumber\\
	{\rm otherwise}&=&0\ .
\label{}
\end{eqnarray}
Note that these equations are not independent. In fact, a derivative of the first equation with respect to $r$ leads the second equation.  
Therefore, it is sufficient to consider only the first equation. 
Integrating this equation leads 
\begin{eqnarray}
	 W[\psi]=\frac{\mu}{r^{n+1}}-\frac{Q^2}{r^{2n}}\equiv M(r)\ .\label{poly}
\end{eqnarray}
In (\ref{poly}),  $\mu$ is constant of integral which is related to ADM mass as 
\begin{eqnarray}
	M_{ADM}=\frac{2\mu\pi^{(n+1)/2}}{\Gamma((n+1)/2)} \ , \label{eq:ADM}
\end{eqnarray}
where we used a unit $16\pi G=1$. We can get this result from the asymptotic behavior~\cite{Myers:1988ze}  and also gain by a
bzackground-independent formalism.~\cite{Kofinas:2007ns}\ 

 In this paper, we want to concentrate on asymptotically flat, i.e. $\Lambda=0$, solutions 
with a positive ADM mass  $\mu>0$
  because such black holes could be created at the LHC. 
We also assume that Lovelock coefficients satisfy
\begin{eqnarray}
	a_m\geq 0 \ , \label{conditions}
\end{eqnarray}
for simplicity.  
Furthermore, for numerical calculation, 
we nondimensionalize all variables. 
Our choice $a_1=1$ means only  scale of length is not fixed. 
There exist many candidates, but  we use $\sqrt{\alpha_2}$ for nondimensionalization in this paper. 
For example, radial $r$ can be nondimensionalized as ${\tilde r}\equiv r/\sqrt{\alpha_2}$. 
From $f(r)=1-r^2\psi(r)=1-{\tilde r}^2(\alpha_2\psi(r))$, $\psi(r)$ should be nondimensionalized as 
${\tilde \psi}\equiv\alpha_2\psi$ and this nondimensionalization leads that eq. (\ref{poly}) is expressed as 
\begin{eqnarray}
{\tilde \psi}+\frac{1}{2}{\tilde \psi}^2+\sum_{m=3}^k \frac{c_m}{m}{\tilde \psi}^m=\frac{{\tilde \mu}}{{\tilde r}^{n+1}}-\frac{{\tilde Q}^2}{{\tilde r}^{2n}}\ ,
\label{poly2}
\end{eqnarray}
where 
\begin{eqnarray}
c_m\equiv\frac{\alpha_m}{\alpha_2^{m-1}},\quad {\tilde \mu}\equiv\alpha_2^{-(n-1)/2}\mu,\quad |{\tilde Q}|\equiv\alpha_2^{-(n-1)/2}|Q|
\label{}
\end{eqnarray}
are nondimensionalized Lovelock coefficients,  nondimensionlized mass parameter and nondimensionalized charge parameter respectivrely. 
After this, we miss tilde, that is,  $r$ means nondimensionlized radius ${\tilde r}$ for example. 
\begin{figure}[t]
 \begin{center}
  \includegraphics[width=80mm]{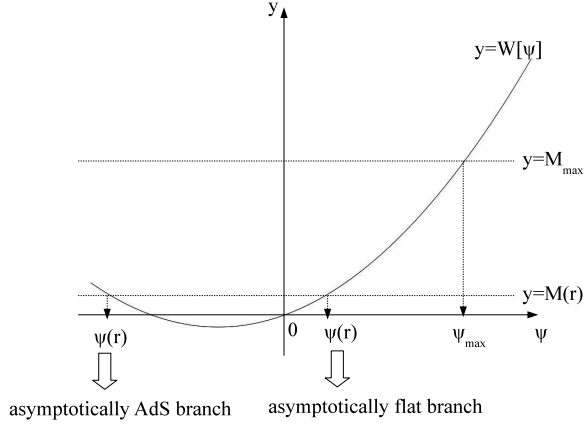}
 \end{center}
 \caption{In this figure, the curve means $y=W[\psi]$ and the dotted line is $y=M(r)$. The cross points of these mean the roots of (\ref{poly2}). 
 Note that radial $r$ is fixed in this figure so $M(r)$ is constant.}
 \label{fig1}
\end{figure}

It is  easily seen that (\ref{poly2}) has many branches. 
However, under our assumptions (\ref{conditions}), an asymptotic flat solution must exist. 
For example, in $k=2$ case, (\ref{poly2}) leads two branches 
\begin{eqnarray}
\psi(r)=
\left\{
\begin{array}{l}
-1+\sqrt{1+2\mu/r^{n+1}-2Q^2/r^{2n}}\\
-1-\sqrt{1+2\mu/r^{n+1}-2Q^2/r^{2n}}
\end{array}
\right.
\ .
\label{k2psi}
\end{eqnarray}
In these two branches, the upper branch is asymptotic flat branch because this behaves as $f(r)=1-r^2\psi(r)\sim 1-\mu/r^{n-1}$ in asymptotic region. 

Against $k=2$ case, it is difficult to solve (\ref{poly2}) in general cases. 
However, we can recognize that there must exist an asymptotic flat branch  
by using graphically method which is expressed in Fig.\ref{fig1}.
In this figure, we fix $r$ and seek the corresponding roots $\psi$, that is, $\psi(r)$. 
In detail, we set the both sides of (\ref{poly2}) as $y$ and draw two curves $y=W[\psi]$ and $y=M(r)(={\rm const.})$ 
in $\psi - y$ plane. The cross points of these curves correspond $\psi(r)$. 
In order to check the existence of asymptotic flat branch, we must take care two points. 
The first point is behavior of $W[\psi]$ in $\psi\geq 0$. Using our assumptions (\ref{conditions}), 
it is easily seen  that $W[0]=0$ and $W[\psi]$ is monotonically increasing in $\psi>0$. 
The second point is behavior of $M(r)$, especially in asymptotic region. 
\begin{figure}[t]
 \begin{center}
  \includegraphics[width=80mm]{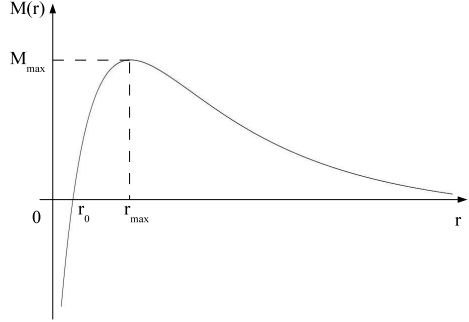}
 \end{center}
 \caption{This figure shows the behavior of $M(r)$. $M(r)$ is always negative in $r<r_0=(Q^2/\mu)^{1/(n-1)}$. 
$M(r)$ is monotonically increasing in $r<r_{max}=(2nQ^2/((n+1)\mu))^{1/(n-1)}$ , 
$M=M_{max}=\mu\frac{n-1}{2n}\left(\frac{(n+1)\mu}{2nQ^2}\right)^{(n+1)/(n-1)}$ at $r=r_{max}$ and 
$M(r)$ is monotonically decreasing in $r>r_{max}$. In asymptotic region $r\rightarrow \infty$, this function approaches $0$}
 \label{fig2}
\end{figure}
This function behaves as Fig.\ref{fig2}. From this figure, 
it is easily seen that $M(r)$ is always positive in asymptotic region and $M(r)\rightarrow 0$ as $r\rightarrow \infty$. 
Combining these two points and Fig.\ref{fig1}, 
we can know that  there must exist a cross point  of $y=W[\psi]$ and $y=M(r)$ in $\psi>0$ as long as we consider asymptotic region; that is, 
a root of (\ref{poly2}) which satisfy $\psi(r)>0$ must exist in asymptotic region.  
And this root also satisfies $\psi\rightarrow 0$ as $r\rightarrow \infty$ because $M(r)\rightarrow 0$ in this limit. 
Considering eq.(\ref{poly2}), $\psi\sim 0$ means $\psi \sim \mu/r^{n+1}-Q^2/r^{2n}$ and $f(r)\sim 1-\mu/r^{n-1}+Q^2/r^{2n-2}$ which means  
this root expresses asymptotic flat. Note that this result also reminds us that the constant of integration $\mu$ is proportional to ADM mass.
After this, we only consider this asymptotic flat branch. 

From Fig.\ref{fig1},  this asymptotic flat branch behaves $0\rightarrow \psi_{max}\rightarrow 0$ as r moves $\infty\rightarrow r_{max}\rightarrow r_0$, 
and $\psi$ becomes negative in $r<r_0$ because $M(r)$ is negative in this region. 
When $r$ becomes still smaller, this branch runs into singularity where Kretschmann invariant   $R_{\mu\nu\rho\sigma}R^{\mu\nu\rho\sigma}$ diverges. 
This variable can be calculated as 
\begin{eqnarray}
	R_{\mu \nu \lambda \rho}R^{\mu \nu \lambda \rho}=f^{''}+2n\frac{f^{'2}}{r^2}+2n(n-1)\frac{(1-f)^2}{r^4}\nonumber
\end{eqnarray}
from metric ansatz (\ref{metric_ansatz}). 
Hence, there are singularities at $r=0$ and also exist where derivatives of $f(r)$ diverge. 
Especially, for the latter,  the first derivative of $f(r)$ is 
\begin{eqnarray}
f^{\prime}=-2r\psi-r^2\psi^{\prime}=-2r\psi-r^2\frac{1}{\partial_{\psi}W[\psi]}\left(-(n+1)\frac{\mu}{r^{n+2}}+2n\frac{Q^2}{r^{2n+1}}\right)\ ,
\label{}
\end{eqnarray}
so Kretschmann invariant diverges at $\partial_{\psi}W[\psi]=0$, that is, 
singularities also exist  where $y=W[\psi]$ becomes extreme value. 
\begin{figure}[t]
  \begin{center}
   \includegraphics[width=100mm]{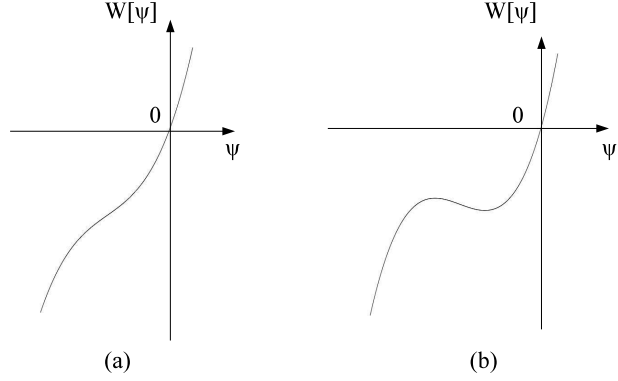}
  \end{center}
  \caption{(a): $W[\psi]$ has no extreme value in $\psi<0$. (b): $W[\psi]$ has extreme values in $\psi<0$.
  Whether $W[\psi]$ has extreme value or not depends on Lovelock coefficients $a_m$ and $k$.} 
  \label{fig3}
  \end{figure}
  This means that position of singularity is different between Fig.\ref{fig3}-(a) and Fig.\ref{fig3}-(b). 
  If $W[\psi]$ has no extreme values in $\psi<0$ (Fig.\ref{fig3}-(a)), there is a singularity at $r=0$: 
  if $W[\psi]$ has extreme values at $\psi(r_s)<0$(Fig.\ref{fig3}-(b)), there is a singularity at $r=r_s$ $(0<r_s<r_0)$. 
  In either case, there is a singularity in $r<r_0$. 

From the viewpoint of cosmic censorship conjecture, there must exist horizons outside of singularities.  
The branch we concentrate on is static and asymptotic flat, so event horizon coincides with killing horizon. 
Then horizon radius $r_H$ is characterized as 
\begin{eqnarray}
0=f(r_H)=1-r_H^2\psi(r_H)\equiv 1-r_H^2\psi_H\ ,
\label{f_0}
\end{eqnarray}
which means $\psi_H$ must be positive. 
This result and behavior of $\psi$ show that $r_H$, if exists, is larger than $r_0$; that is, 
there exists no naked singularity if $f(r_H)=0$ has solutions. 
Hence, what we have to examine is conditions for existence of horizons. 
Horizons are characterized by (\ref{f_0}) and $\psi_H$ also satisfies 
\begin{eqnarray}
W[\psi_H]=M(r_H)\ .\nonumber
\label{}
\end{eqnarray}
Therefore horizons are determined by two equations
\begin{eqnarray}
\left\{
\begin{array}{l}
\psi_H=\frac{1}{r_H^2}\\
W[\psi_H]=\frac{\mu}{r_H^{n+1}}-\frac{Q^2}{r_H^{2n}}
\end{array}
\right.\ .
\label{}
\end{eqnarray}
\begin{figure}[t]
  \begin{center}
   \includegraphics[width=70mm]{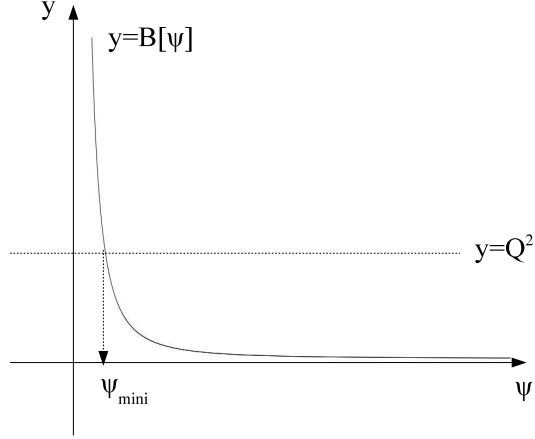}
  \end{center}
  \caption{In this figure, we draw two lines $y=B[\psi]$ and $y=Q^2$ in $\psi-y$ plane. The cross point means $\psi_{mini}$.
  Under our assumption (\ref{conditions}), $y=B[\psi]$ is monotonic function which diverges near $\psi \sim 0$ and approaches $0$  as $\psi \rightarrow \infty$.}
  \label{fig4}
  \end{figure}
Substituting the first equation for the second equation leads  
 \begin{eqnarray}
&\ &W[\psi_H]=\mu \psi_H^{(n+1)/2}-Q^2\psi_H^{n}\nonumber\\
&\Leftrightarrow&\mu=Q^2\psi_H^{(n-1)/2}+\psi_H^{-(n+1)/2}W[\psi_H]\equiv A[\psi_H]\ .
\label{det_horizon}
\end{eqnarray} 
The solutions of this equation correspond horizon, 
so the behavior of $y=A[\psi]$ determines the condition for existence of horizons. 
The derivative of $A[\psi]$ with respect to $\psi$ is 
\begin{eqnarray}
\partial_{\psi}A[\psi]=\frac{n-1}{2}Q^2\psi^{(n-3)/2}-\frac{n+1}{2}\psi^{-(n+3)/2}W[\psi]+\psi^{-(n+1)/2}\partial_{\psi}W[\psi]\label{d_of_A}
\end{eqnarray}
and solving $\partial_{\psi}A[\psi]=0$ with respect to $Q^2$ leads 
\begin{eqnarray}
Q^2&=&\frac{2}{n-1}\left[\frac{n+1}{2}\psi^{-n}W[\psi]-\psi^{-n+1}\partial_{\psi}W[\psi]\right]\nonumber\\
&=&\frac{2}{n-1}\left[\left(\frac{n+1}{2}-1\right)\psi^{1-n}+\left(\frac{n+1}{4}-1\right)\psi^{2-n}+\sum_{m=3}^k\frac{c_m}{m}\left(\frac{n+1}{2}-m\right)\psi^{m-n}\right]\nonumber\\
&\equiv&B[\psi]\ .
\label{kyokusyou}
\end{eqnarray}
Note that $n=2k-1$ or $n=2k$. Then 
powers of all terms in $B[\psi]$ are negative and their coefficients are positive. 
Therefore, $y=B[\psi]$ is monotonically decreasing function like Fig.\ref{fig4}.  
From this figure, there must exist a root of (\ref{kyokusyou}). We call this $\psi_{mini}$ because 
$A[\psi]$ becomes extreme minimum at $\psi_{mini}$ which  can be seen from (\ref{d_of_A}). 
\begin{figure}[t]
  \begin{center}
   \includegraphics[width=80mm]{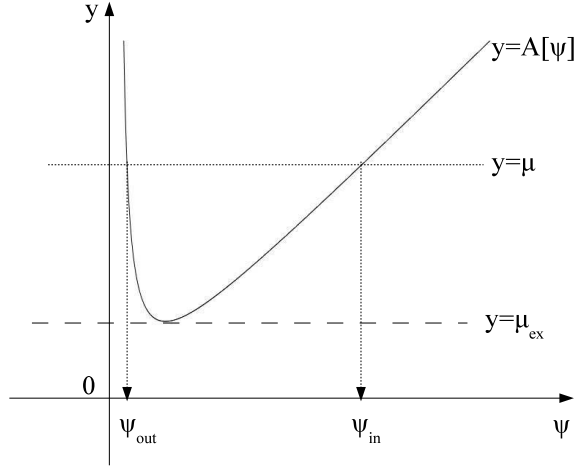}
  \end{center}
  \caption{In this figure, $y=A[\psi]$ and $y=\mu$ are drawn. The cross points of these lines correspond $\psi_H$. 
  From this figure, it can be easily seen that there exist horizons if $\mu\geq \mu_{ex}$.}
  \label{fig5}
  \end{figure}
  Then, $A[\psi]$ behaves as Fig.\ref{fig5}. In this figure, we solve eq.(\ref{det_horizon}) graphically.
  The cross points of $y=\mu$ and $y=A[\psi]$ mean horizons. From this figure, it can be easily seen that 
  there exist horizons if $\mu\geq \mu_{ex}$ where $\mu_{ex}$ is defined as 
\begin{eqnarray}
\mu_{ex}=Q^2\psi_{mini}^{(n-1)/2}+\psi_{mini}^{-(m+1)/2}W[\psi_{mini}]\ .
\label{mu_ex}
\end{eqnarray}
Note that $\psi_{out}$ and $\psi_{in}$ in Fig.\ref{fig5} means outer horizon and inner horizon respectively 
because $\psi_{H}$ corresponds horizon radius as $\psi_H=1/r_H^2$. 

Finally, for example, we examine $\mu_{ex}$ in 5-dimensions ($n=3$ and $k=2$ case). 
From eq.(\ref{kyokusyou}), $\psi_{mini}$ in 5-dimensions is 
\begin{eqnarray}
Q^2=\frac{1}{\psi_{mini}^2}\Leftrightarrow \psi_{mini}=\frac{1}{|Q|}\ ,\nonumber
\label{}
\end{eqnarray}
so substituting this for (\ref{mu_ex}) leads 
\begin{eqnarray}
\mu_{ex}=2|Q|+\frac{1}{2}\ .
\label{mu_ex_5}
\end{eqnarray}

 
\section{Condition for Stability under Tensor Type Perturbations}
\label{seq:3}
In this section, we consider conditions for stability under tensor type perturbations. 
Note that we only consider $\mu\geq \mu_{ex}$ which means there exists horizon and 
we only consider the perturbation outside of the outer horizon. 

The background metric (\ref{metric_ansatz}) has $n$ dimensional spherical symmetry, so 
perturbations are decomposed into scalar type, vector type and tensor type. 
In this paper, we concentrate on tensor type perturbations which are characterized as 
\begin{eqnarray}
\delta g_{tt}&=&\delta g_{tr}=\delta g_{ti}=\delta g_{rr}=\delta g_{ri}=0\nonumber\\
\delta g_{ij}&=&r^2\chi(r)e^{i\omega t}h_{ij}^{T}\ ,
\label{}
\end{eqnarray}
where $\chi$ means master variable and $h_{ij}^{T}$ is tensor harmonics. 
Tensor harmonics $h_{ij}^{T}$ satisfies $\gamma^{ij}h_{ij}^T=0$, $h^T_{ij}{}^{|j}=0$ and $h^T_{ij}{}^{|k}{}_{|k}=-\gamma_t h_{ij}^T$ 
where $\gamma_t=\ell(\ell+n-1)-2$ $(\ell=2,3,4,\cdots)$. Here $|$ means covariant derivative with respect to $S^n$ metric $\gamma_{ij}$.
Note that there is no tensorial perturbation for Maxwell field $A_{\mu}$, which means that the first order perturbation of 
energy momentum tensor $\delta T_{\mu}{}^{\nu}=0$. 
Therefore, the first order perturbation of EOM is $\delta{\cal G}_{\mu}{}^{\nu}=0$ and 
this can be calculated as follows;\ \cite{Takahashi:2009dz}\ 
\begin{eqnarray}
- f^2T^{'}\chi^{''}
      - \left( f^2 T^{''}+ \frac{2f^2T^{'}}{r}
      +  f f^{'}T^{'} \right) \chi^{'}
      +  \frac{\ell(\ell+n-1)f}{(n-2)r}T^{''} \chi 
      =   \omega^2 T^{'}\chi\ .
\label{tensor_EOM}
\end{eqnarray}
Here $T(r)$ is defined as 
\begin{eqnarray}
T(r)=r^{n-1}\partial_{\psi}W[\psi]\ .
\label{}
\end{eqnarray}

Now we will present the condition for the stability of the solutions we are considering in this paper. 
   
As we will soon see, the master equation (\ref{tensor_EOM}) can be transformed
 into the Schr${\rm {\ddot o}}$dinger form. 
To do this, we have to impose the condition 
\begin{eqnarray}
	T^{'}(r)>0  \quad ({\rm for}\ r>r_H) \ . \label{eq:assum}
\end{eqnarray}
In fact, this is necessary for the linear analysis to be applicable. 
 In the case that there exists $r_{g}$ such that $T^{'}(r_g)=0$ and $r_g>r_H$,
 we encounter a singularity. 
 Using approximations $T^{'}(r)\sim T^{''} (r_g)(r-r_g)\equiv T^{''}(r_g) y$, 
$f(r)= f(r_g) $ and $r=r_g$, (\ref{tensor_EOM}) approximately becomes 
\begin{eqnarray}
	y \frac{d^2\chi}{dy^2}+\frac{d\chi}{dy}+c \chi=0  \ .
\end{eqnarray}
This  shows that near $r=r_g$, $\chi$ behaves as  $\chi \sim c_1 + c_2 \log y$ , 
where $c_1$ and $c_2$ are constants of integration. 
Hence, the solution is singular at $y =0$ for generic perturbations. 
 The similar situation occurs even in cosmology with higher derivative
  terms.\cite{Kawai:1998bn,Satoh:2007gn} \ 
 In those cases, this kind of singularity alludes to ghosts.
 Indeed, if there is a region $T^{'}(r)<0$ outside the horizon, the kinetic
 term of perturbations has a wrong sign.
 Hereafter, we call this the ghost instability. 

When the condition (\ref{eq:assum}) is fulfilled, introducing a new variable
$
	\Psi(r)=\chi(r)r\sqrt{T^{'}(r)} 
$
 and switching to the coordinate $r^*$, defined by $dr^*/dr=1/f$,
we can rewrite Eq.(\ref{tensor_EOM}) as
\begin{eqnarray}
	-\frac{d^2\Psi}{dr^{*2}}+V_t(r(r^*))\Psi=\omega^2\Psi \ , 
      \label{eq:shradinger}
\end{eqnarray}
where
\begin{eqnarray}
	V_t(r)=\frac{\ell(\ell+n-1)f}{(n-2)r}\frac{d \ln{T^{'}}}{dr}+\frac{1}{r\sqrt{T^{'}}}f\frac{d}{dr}\left(f\frac{d}{dr}r\sqrt{T^{'}}\right) \label{eq:potential}
\end{eqnarray}
is an effective potential. 

For discussing the stability, the ``S-deformation" approach
 is useful.\cite{Kodama:2003jz, Dotti:2004sh} \ 
Let us define the operator 
\begin{eqnarray}
	{\cal H}\equiv -\frac{d^2}{dr^{*2}}+V_t
\end{eqnarray}
acting on smooth functions defined on $I=(r^{*}_H,\infty)$.
Then, (\ref{eq:shradinger}) is the eigenequation and $\omega^2$ is eigenvalue 
of ${\cal H}$. 
We also define the inner products as
\begin{eqnarray}
	(\varphi_1,\varphi_2)=\int_I \varphi_1^*\varphi_2 dr^* \ .\label{inner_product}
\end{eqnarray}

In this case, for any smooth function $\varphi$ with compact support in $I$, we can find a smooth function $S$ such that 
\begin{eqnarray}
	(\varphi,{\cal H}\varphi)=\int_{I} (|D\varphi|^2+\tilde{V}|\varphi|^2)dr^{*},
\end{eqnarray}
where we have defined
\begin{eqnarray}
	D=\frac{d}{dr^{*}}+S  \ , \quad 
	\tilde{V}=V_t+f\frac{dS}{dr}-S^2  \ .\label{tildeV}
\end{eqnarray}
Following Ref.~\citen{Dotti:2004sh}, we choose $S$ to be
\begin{eqnarray}
	S=-f\frac{d}{dr}\ln{(r\sqrt{T^{'}})} \ .
\end{eqnarray}
Then, we obtain the formula
\begin{eqnarray}
	(\varphi,{\cal H}\varphi)
      =\int_{I} |D\varphi|^2dr^{*}+\ell(\ell+n-1)
      \int_{r_H}^{\infty}\frac{|\varphi|^2}{(n-2)r}\frac{d \ln{T^{'}}}{dr}dr \ .
       \label{eq:stab}
\end{eqnarray}
Here, the point is that the second term in (\ref{eq:stab}) includes
a factor $\ell(\ell+n-1)>0$, but $T^{'}$ does not include $\ell$. 
Hence, by taking a sufficiently large $\ell$, we can 
always make the second term dominant.  

Now, let us show that the sign of $d\ln T^{'} /dr$ determines the stability.
If $d\ln T^{'} /dr >0$ on $I$, the solution (\ref{metric_ansatz}) is stable.
This can be understood as follows. 
Note that $\ell(\ell+n-1)>0$, 
then we have $\tilde{V}>0$ for this case.
 That means $(\varphi,{\cal H}\varphi)>0$ for arbitrary $\varphi$ if $d \ln{T^{'}}/dr>0$ on $I$. 
We choose, for example, $\varphi$ as the lowest eigenstate, then we can conclude that the lowest eigenvalue $\omega^2_0$ is positive. 
Thus, we proved the stability. The other way around,
if $d \ln{T^{'}} /dr <0$ at some point in $I$, the solution is unstable. 
To prove this, the inequality
\begin{eqnarray}
	\frac{(\varphi,{\cal H}\varphi)}{(\varphi,\varphi)} \geq \omega_0^2
      \label{eq:ineq}
\end{eqnarray}
is useful. 
This inequality is correct for arbitrary $\varphi$, which is not necessarily an eigenfunction of ${\cal H}$, as far as the left hand side of Eq. (\ref{eq:ineq}) make sense.
 If $d \ln{T'}/dr <0$ at some point in $I$, we can find $\varphi$ such that
\begin{eqnarray}
	 \int_{r_H}^{\infty}\frac{|\varphi|^2}{(n-2)r}\frac{d \ln{T^{'}}}{dr}dr<0 \ .
\end{eqnarray}
In this case, (\ref{eq:stab}) is negative for sufficiently large $\ell$.
 Then, the inequality (\ref{eq:ineq}) implies $\omega^2_0<0$ 
 and the solution has unstable modes. 
Thus, we can conclude that the solution is stable if and only if
 $d \ln{T^{'}}/dr>0$ on $I$.
 
 From the above logic, if $d\ln{T^{'}}/dr$ has a negative region, 
 negative $\omega^2$ states exist. 
 Therefore, this instability is dynamical. Then, we call this as dynamical instability in order to distinguish this from the 
 ghost instability which is caused by negativity of $T^{'}(r)$. 
 
We want to summarize this section. 
If $T^{'}$ has negative region outside the outer horizon $r>r_{out}$, this solution
has the ghost instability. Even if $T^{'}$ is always positive, 
this solution has the dynamical instability 
if $T^{''}$ has a negative region outside the outer horizon. 
Therefore, charged Lovelock black holes are stable under tensor perturbations 
if and only if $T^{'}$ and $T^{''}$ are always positive outside the outer horizon. 

 Note that $T$ is calculated as $T=r^{n-1}\sqrt{1+2M(r)}$ in $k=2$ case from (\ref{k2psi}). 
 This leads 
 \begin{eqnarray}
T^{\prime}=\frac{r^{n-2}}{\sqrt{1+2M(r)}}\left[(n-1)+(n-3)\frac{\mu}{r^{n+1}}+2\frac{Q^2}{r^{2n}}\right]>0, 
\label{}
\end{eqnarray}
which means there exist no ghost instabilities in 5 and 6 dimensions. 

\section{Numerical Results} 
\label{seq:4}
In this section, we examine ``whether $T^{\prime}$ has negative region or not" and ``whether $T^{\prime\prime}$ has negative region or not" numerically. 
The detail is as follows. Note that $\Delta \mu, d\mu, d|Q|$ and $Q_{end}$ are parameters for numerical calculations. 

Firstly, we fix $|Q|$. Using this $|Q|$,  extreme mass $\mu_{ex}$ can be determined from (\ref{mu_ex}). 
Then, we change $\mu$ from $\mu_{ex}$ to $\mu_{ex}+\Delta \mu$ by $d\mu$. 
For each $\mu$, we can gain $\psi(r)$ with $(|Q|,\mu)$ numerically from (\ref{poly2}) and  determine $T^{\prime}(r)$ and $T^{\prime\prime}(r)$. 
Then, checking  whether  these functions have negative region or not, 
we determine, for example, the border between stable and unstable which we call  $\mu_{max}$. 
Then, we change $|Q|$ by $d|Q|$ and do the same calculation.  
We repeat this calculation until $|Q_{end}|$. 

Note that we examine the region $r_{out}\leq r\leq r_{out}+10$ and 
the mesh size $dr$ is $dr=1.0\times 10^{-3}$. 

\subsection{5-dimensional Case}

\begin{figure}[htbp]
  \begin{center}
   \includegraphics[width=90mm]{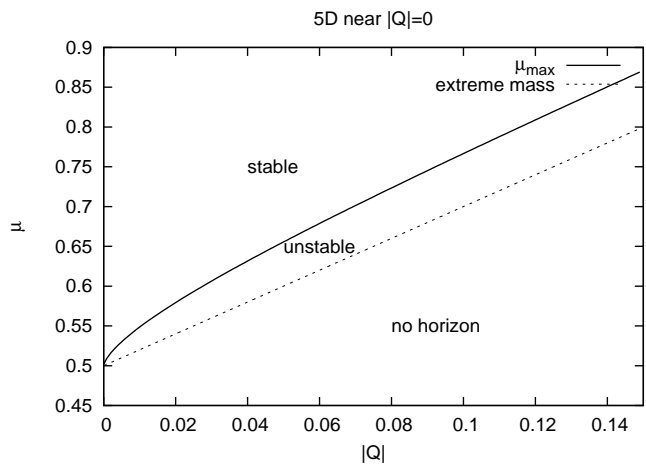}
  \end{center}
  \caption{$Q-\mu$ diagram near $|Q|=0$ in 5-dimensions. This is calculated with $\Delta \mu=0.1$ and $\ d\mu=1.0\times 10^{-4}$. And $d|Q|$
  is $d|Q|=1.0\times 10^{-4}$ when $|Q|<10^{-3}$ and $d|Q|=1.0\times 10^{-3}$ in $|Q|>10^{-3}$. 
  The linearity of extreme mass line is guaranteed by (\ref{mu_ex_5})}
  \label{fig6}
  \end{figure}

\begin{figure}
 
 \begin{center}
  \includegraphics[width=90mm]{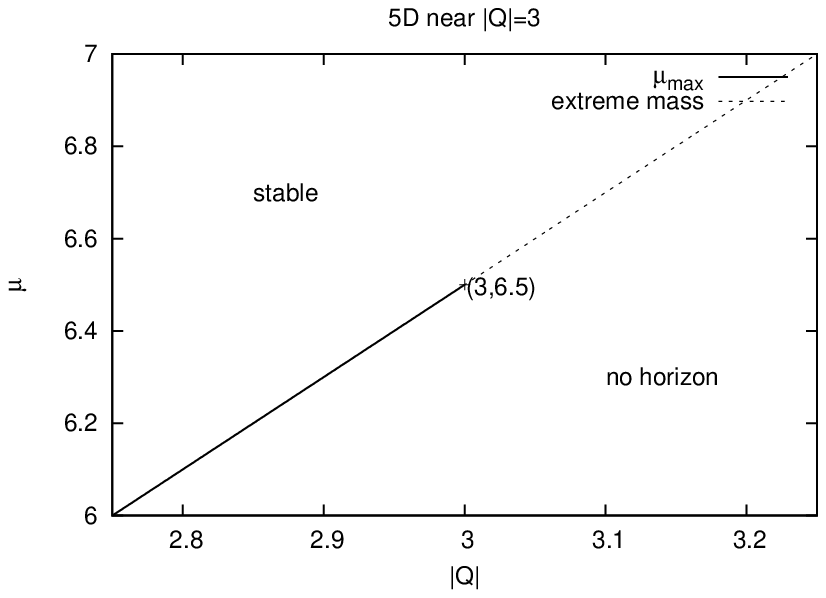}
 \end{center}
  \caption{$Q-\mu$ diagram near $|Q|=3$ in 5-dimensions. This is calculated with $\Delta \mu=0.05$ and $\ d\mu=d|Q|=1.0\times 10^{-3}$. 
  In this figure, the difference between $\mu_{max}$ and extreme mass is  $O(10^{-3})$ where $|Q|\leq3$  and there is no instability where $|Q|>3$. }
  \label{fig7}
\end{figure}

Numerical results of 5-dimensional case are Fig.\ref{fig6} and Fig.\ref{fig7}. 
The former is $Q-\mu$ diagram near $|Q|\sim 0$ and the latter is that of $|Q|\sim 3$. 
It can be seen that black holes with  $|Q|=0$ are stable, 
which agrees with our previous results.~\cite{Takahashi:2009dz}\  
When black holes are a little charged up, however, there exists an unstable region near extreme mass (Fig.\ref{fig6})
and this region vanishes in $|Q|\geq 3$ (Fig.\ref{fig7}). 
As we have already discussed, there is no ghost instability in 5-dimensions.  
Hence, these figures show that  nearly extreme black holes are unstable if $0<|Q|\leq 3$. 
\subsection{6-dimensional Case}

\begin{figure}[htbp]
 \begin{center}
  \includegraphics[width=90mm]{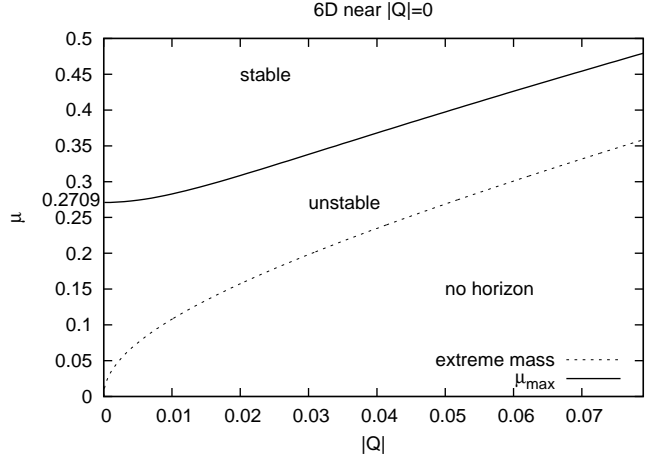}
 \end{center}
 \caption{$Q-\mu$ diagram near $|Q|=0$ in 6-dimensions. This is calculated with $\Delta \mu=0.05$ and $\ d\mu=1.0\times 10^{-4}$. And $d|Q|$
  is $d|Q|=1.0\times 10^{-4}$ when $|Q|<10^{-3}$ and $d|Q|=1.0\times 10^{-3}$ in $|Q|>10^{-3}$.}
 \label{fig8}
\end{figure}

\begin{figure}[htbp]
 \begin{center}
  \includegraphics[width=90mm]{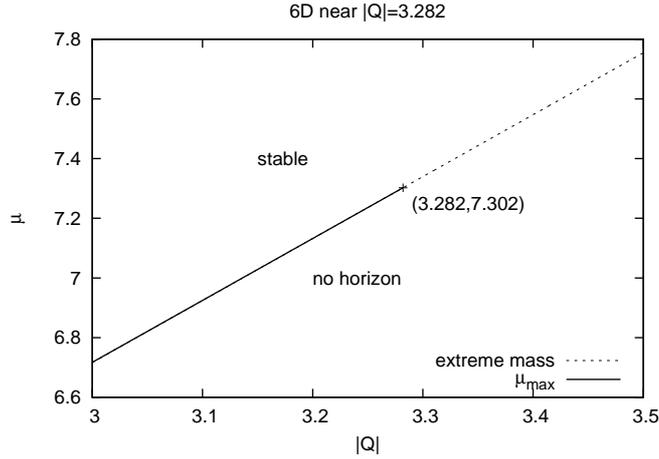}
 \end{center}
 \caption{$Q-\mu$ diagram near $|Q|=3.28$ in 6-dimensions. This is calculated with $\Delta \mu=0.01$ and $d\mu=d|Q|=1.0\times 10^{-3}$.
In this figure, the difference between $\mu_{max}$ and extreme mass is  $O(10^{-3})$ in $|Q|\leq 3.282$  and there is no instability in $|Q|>3.282$. }
 \label{fig9}
\end{figure}

Fig.\ref{fig8} and Fig.\ref{fig9} are the numerical results in 6-dimensions. 
The former is $Q-\mu$ diagram near  $|Q|=0$ and the latter is that of $|Q|\sim 3.28$. 
When black holes are neutral, there is unstable region in $0<\mu\leq 0.27$; this agrees with 
our previous results.~\cite{Takahashi:2009dz}\  
When black holes are a little charged up, there also exist an unstable region (Fig.\ref{fig8}). 
However, this region vanishes in $|Q|>3.282$ (Fig.\ref{fig9}). 
As we have already discussed, there is no ghost instability in 6-dimensions.
Therefore, it is shown numerically that nearly extreme black hole is unstable when its charge satisfies $0<|Q|\leq 3.282$. 

\subsection{7-dimensional case}

The previous works tell that neutral black holes do not have ghost instability if $c_3<0.25$ and have this instability if $c_3>0.25$.
Then, in this paper, we concentrate on $c_3=0.2$ case and $c_3=0.3$ case. 

\subsubsection{$c_3=0.2$ case}

\begin{figure}[htbp]
 \begin{center}
  \includegraphics[width=90mm]{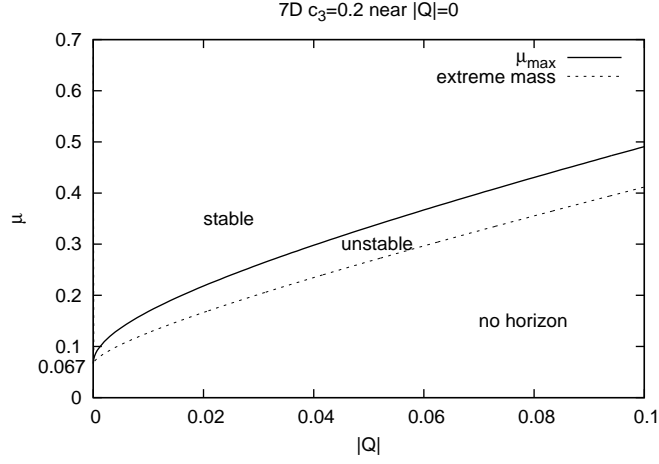}
 \end{center}
 \caption{$Q-\mu$ diagram near $|Q|=0$ when $c_3=0.2$ in 7-dimensions. This is calculated with $\Delta \mu=0.15$ and $\ d\mu=1.0\times 10^{-4}$. And $d|Q|$
  is $d|Q|=1.0\times 10^{-4}$ when $|Q|<10^{-3}$ and $d|Q|=1.0\times 10^{-3}$ in $|Q|>10^{-3}$.}
 \label{fig10}
\end{figure}

\begin{figure}[htbp]
 \begin{center}
  \includegraphics[width=90mm]{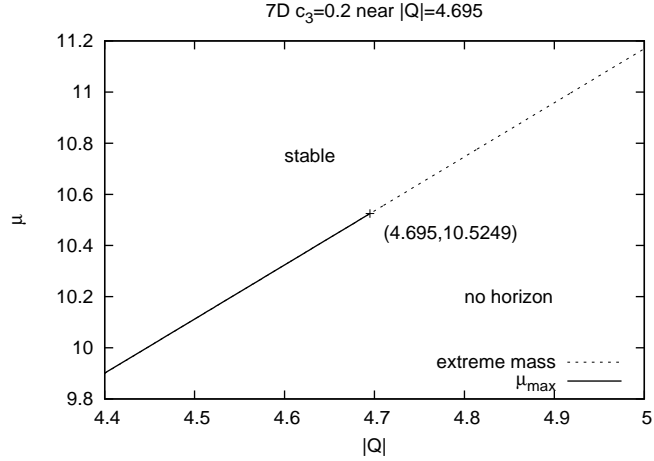}
 \end{center}
 \caption{$Q-\mu$ diagram near $|Q|=4.7$ when $c_3=0.2$ in 7-dimensions. This is calculated with $\Delta \mu=1.0\times 10^{-3}$, $d\mu=1.0\times 10^{-4}$ and $d|Q|=1.0\times 10^{-3}$.
In this figure, the difference between $\mu_{max}$ and extreme mass is  $O(10^{-4})$ in $|Q|\leq 4.695$  and there is no instability in $|Q|>4.695$. }
 \label{fig11}
\end{figure}

Fig.\ref{fig10} and Fig.\ref{fig11} are the numerical results when $c_3=0.2$ in 7-dimension. 
The former is $Q-\mu$ diagram near  $|Q|=0$ and the latter is that of $|Q|\sim 4.7$. 
In Fig.\ref{fig10}, there are no unstable region and  no ghost region, which are consistent with our previous results.~\cite{Takahashi:2009dz}\ 
However, when a little charged up, black hole has dynimical instability if its mass is as small as extreme mass. 
And this instability vanishes when $|Q|>4.695$ (Fig.\ref{fig11}). 
Note that ghost region cannot be detected by our numerical calculation with $0\leq |Q|\leq 5$, $\Delta \mu$=1 and $d\mu=d|Q|=0.01$. 
Therefore, same as 5-dimensional case, these result means that there exists an unstable region near extreme mass in $0<|Q|\leq 4.695$.   

\subsubsection{$c_3=0.3$ case}

\begin{figure}[htbp]
 \begin{center}
  \includegraphics[width=90mm]{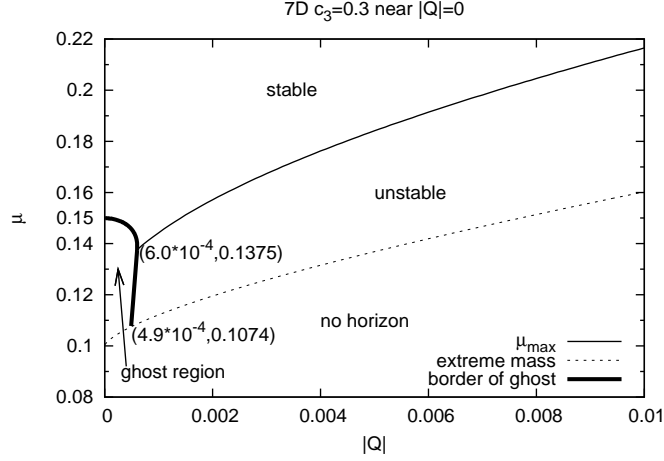}
 \end{center}
 \caption{$Q-\mu$ diagram near $|Q|=0$ when $c_3=0.3$ in 7-dimensions. $\mu_{max}$ is calculated with $\Delta \mu=0.1$ and $\ d\mu=1.0\times 10^{-4}$. 
 Ghost and extreme mass are done with $\Delta \mu=0.1,\ d\mu=1.0\times 10^{-4}$ and $d|Q|=1.0\times 10^{-5}$. 
 This figure shows that there exists a ghost region if $|Q|<6.0\times 10^{-4}$ and there exists an unstable region if $|Q|>4.9\times 10^{-4}$. }
 \label{fig12}
\end{figure}

\begin{figure}[htbp]
 \begin{center}
  \includegraphics[width=90mm]{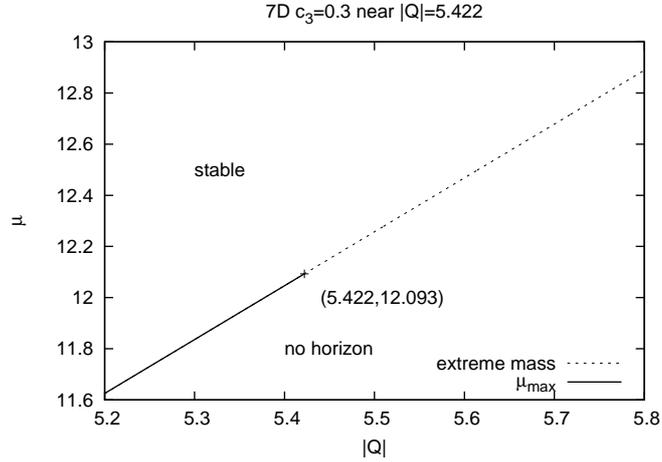}
 \end{center}
 \caption{$Q-\mu$ diagram near $|Q|=5.4$ when $c_3=0.3$in 7-dimensions. This is calculated with $\Delta \mu=1.0\times 10^{-3}$, $d\mu=1.0\times 10^{-4}$ and $d|Q|=1.0\times 10^{-3}$.
In this figure, the difference between $\mu_{max}$ and extreme mass is  $O(10^{-4})$ in $|Q|\leq 5.422$  and there is no instability in $|Q|>5.422$. }
 \label{fig13}
\end{figure}

Fig.\ref{fig12} and Fig.\ref{fig13} are  numerical results when $c_3=0.3$ in 7-dimensions. 
The former is $Q-\mu$ diagram near  $|Q|=0$ and the latter is that of $|Q|\sim 5.4$. 
In Fig.\ref{fig12}, there exists a ghost region if $|Q|=0$ and $0.1<\mu\leq 0.15$. 
This result agrees with the previous work \cite{Takahashi:2009dz}. 
As black hoes are charged up, this ghost region diminishes and unstable region appears. 
And this unstable region vanishes if $|Q|>5.422$ (Fig.\ref{fig13}). 

\subsection{8-dimensional Case}

If black holes are neutral, in 8-dimensions, the previous works show that there is no ghost instability when $c_3<5.92$ and there exists if $c_3>5.92$. 
Therefore, we concentrate on $c_3=5.9$ case and $c_3=6.0$ case in this paper. 

\subsubsection{$c_3=5.9$ case}

\begin{figure}[htbp]
 \begin{center}
  \includegraphics[width=90mm]{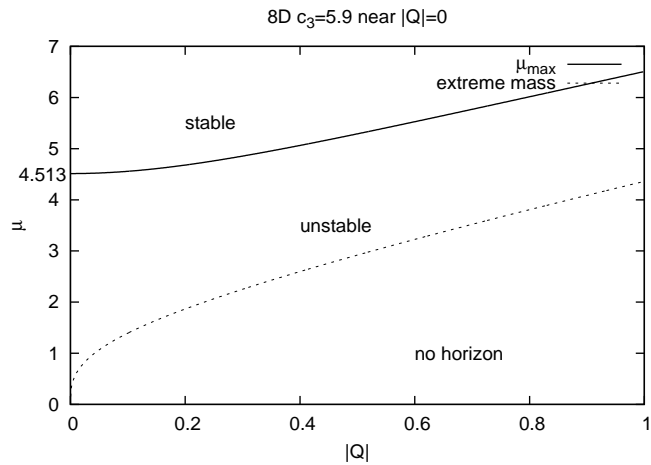}
 \end{center}
 \caption{$Q-\mu$ diagram near $|Q|=0$ when $c_3=5.9$ in 8-dimensions. This is calculated with $\Delta \mu=5$ and $d\mu=d|Q|=1.0\times 10^{-3}$. }
 \label{fig14}
\end{figure}

\begin{figure}[htbp]
 \begin{center}
  \includegraphics[width=90mm]{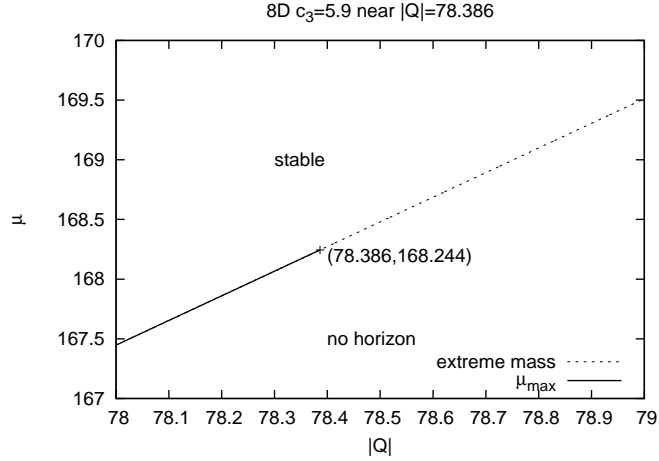}
 \end{center}
 \caption{$Q-\mu$ diagram near $|Q|=78.4$ when $c_3=5.9$ in 8-dimensions. This is calculated with $\Delta \mu=0.01$ and $d\mu=d|Q|=1.0\times 10^{-3}$.
In this figure, the difference between $\mu_{max}$ and extreme mass is  $O(10^{-3})$ in $|Q|\leq 78.386$  and there is no instability in $|Q|>78.386$. }
 \label{fig15}
\end{figure}

Fig.\ref{fig14} and Fig.\ref{fig15} are the numerical results when $c_3=5.9$ in 8-dimension. 
The former is $Q-\mu$ diagram near  $|Q|=0$ and the latter is that of $|Q|\sim 78.4$. 
In Fig.\ref{fig14}, black holes with $|Q|=0$ and $0<\mu<4.513$ are unstable, which agrees with our previous works.~\cite{Takahashi:2009dz}\ 
When a little charged up, black hole has also instability if its mass is as small as extreme mass. 
However this instability vanishes when $|Q|>78.386$ (Fig.\ref{fig15}). 
Note that ghost cannot be found in our numerical calculation with $0\leq |Q|\leq 80$, $\Delta \mu$=5 and $d\mu=d|Q|=0.1$. 
Therefore, same as 6-dimensional case, these result means that there exists an unstable region near extreme mass in $0\leq|Q|\leq 78.386$.   

\newpage
\subsubsection{$c_3=6$ case}

\begin{figure}[htbp]
 \begin{center}
  \includegraphics[width=90mm]{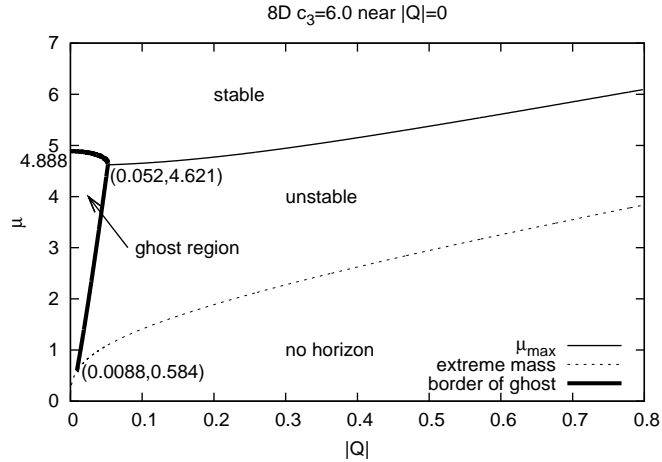}
 \end{center}
 \caption{$Q-\mu$ diagram near $|Q|=0$ when $c_3=6$ in 8-dimensions. $\mu_{max}$ is calculated with $\Delta \mu=5.0$ and $d\mu=d|Q|=1.0\times 10^{-3}$. 
 Ghost and extreme mass are done with $\Delta \mu=5.0,\ d\mu=1.0\times 10^{-3}$ and $d|Q|=1.0\times 10^{-4}$. 
 This figure shows that there exists a ghost region if $|Q|<5.2\times 10^{-2}$ and there exists an unstable region if $|Q|>8.8\times 10^{-3}$. }
 \label{fig16}
\end{figure}

\begin{figure}[htbp]
 \begin{center}
  \includegraphics[width=90mm]{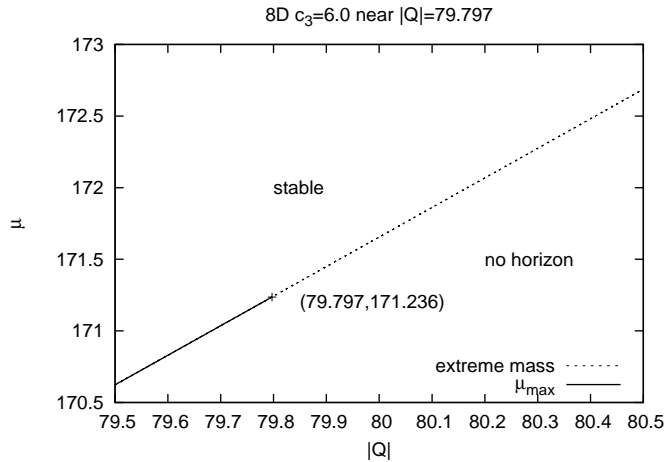}
 \end{center}
 \caption{$Q-\mu$ diagram near $|Q|=79.8$ when $c_3=6$in 8-dimensions. This is calculated with $\Delta \mu=0.01$ and $d\mu=d|Q|=1.0\times 10^{-3}$.
In this figure, the difference between $\mu_{max}$ and extreme mass is  $O(10^{-3})$ in $|Q|\leq 79.797$  and there is no instability in $|Q|>79.797$.}
 \label{fig17}
\end{figure}

Fig.\ref{fig16} and Fig.\ref{fig17} are numerical results when $c_3=6$ in 8-dimensions. 
The former is $Q-\mu$ diagram near $|Q|=0$ and the latter is that of $|Q|\sim 79.8$. 
In Fig.\ref{fig16}, there exists a ghost region if $|Q|=0$ and $0<\mu\leq 4.888$. 
This result agrees with our previous work \cite{Takahashi:2009dz}. 
As black hoes are charged up, this ghost region diminishes and unstable region appears. 
And this unstable region vanishes if $|Q|>79.797$(Fig.\ref{fig17}).

\subsection{Summarizing Numerical Results and Consideration}

In this subsection, we summarize features of  these numerical results and 
consider black hoes at the LHC briefly. 
After this, we call both ``ghost instability" and ``dynamical instability" as only ``instability". 

In neutral case, there exists instability near $\mu=0$ if the spacetime is even dimensions and 
no instability in odd dimensions  with small $c_3$, which agree with the previous results. 
However, a little charged up, black hole suffers instability if its mass is near extreme mass in both even and odd dimensions, 
which is independent of Lovelock coefficient $c_3$.  
Then, we can denote that charge is crucial for Lovelock black hole in odd dimensions. 
Another feature is existence of  $Q_{max}$ such that there is unstable region near extreme mass in $|Q|<Q_{max}$ 
and it vanishes in $|Q|>Q_{max}$. 
This shows that extreme Lovelock black hole must not exist if its charge is less than $Q_{max}$. 

Using  these results, we now consider charged Lovelock black holes at the LHC briefly. 
Because unstable black holes are never created, these events must occur on stable region in $Q-\mu$ diagram. 
So the numerical results show that mass has lower bound $\mu_{max}$ in black hole creation if $|Q|<Q_{max}$.  
Or if we could observe a zero temperature black hole at the LHC, such black hole should has more charge than $Q_{max}$.

After creation, black hole loses its charge quickly by Schwinger-type mechanism \cite{Gibbons:1975kk} 
and ran into evaporating phase.  
It is known that charge of black hole fluctuates in this phase. \cite{Page:1976df}\ 
As we have mentioned, black hole has instability if it has non-zero charge. 
Then, if dimension is  even, black holes always run into unstable region;  
even if dimension is odd, black holes could become unstable by these fluctuations in this phase. 
That is, independent of dimension, black hole has always possibility to become unstable due to fluctuating charge in Lovelock theory while they are evaporating.

\section{Conclusion}
\label{seq:5}

We have studied the stability of static charged black holes under tensor perturbations 
in the second order and the third order Lovelock theory. 
We have derived master equation and presented the criteria for stability. 
Examining these criteria numerically, 
we have shown that there exists  an unstable region near extreme mass in both even and odd dimensions. 
In odd dimensions, especially,  we have already shown that 
there is no dynamical instability when black holes are neutral. 
Therefore, charges have proved fatal to Lovelock black holes especially in odd dimensions. 

We have also shown that this instability is stronger as $\ell$ becomes larger. 
And we have also pointed out that black holes have possibility to ran into 
the unstable region due to fluctuating charges in evaporating phase at the LHC. 
These suggest that black holes in Lovelock theory should be doomed to ruin at the LHC.  

One of future works is extension to vector and scalar type perturbations. 
For vector type perturbations, especially, we have already shown that  black holes are stable in neutral case. 
Therefore, by checking vector type perturbations, 
it may become clearer that charges are perils for Lovelock black holes at the LHC.

Extension to still higher dimensional case is also important. 
This is  because string theory is consistently formulated only in 10-dimensions. 
Although the criterion for stability we have shown in this paper are true for any order Lovelock theory, 
this work is a little bother only because other parameters $c_m$ are added.   

It is interesting to  examine the nature of this instability. 
The shortest way may be search for  the meaning of the function $T(r)$ 
because this governs the dynamical stability of Lovelock black holes. 
Therefore, if $T(r)$ has thermodynamical meaning, the relation between 
thermodynamical~\cite{Cai:2001dz, Dehghani:2005vh} and dynamical instability might be revealed. 
Another way is considering physical properties.  
In Lovelock theory, gravitational collapses have also been studied.~\cite{Maeda:2005ci}\  
They have  shown that apparent horizon is less liable to appear in Lovelock theory, which means 
attractive force becomes weaker; that is, higher curvature corrections are repulsive effectively. 
And this effective repulsive force should be dominant in large $\ell$ mode. 
Then, origin of instability may be  that repulsive force defeats attractive one, 
which is consistent with the results of this paper. 
This repulsive instability might exist in Einstein theory with galileons, too.~\cite{VanAcoleyen:2011mj}\  
\section*{Acknowledgements} 
The author  would like to thank Jiro Soda for his support and encouragement. 
The author also thanks Seiju Ohashi for fruitful discussions. 
This work was supported by the Grant-in-Aid for the Global COE Program 
``The Next Generation of Physics, Spun from Universality and Emergence" 
from the Ministry of Education, Culture, Sports, Science and Technology (MEXT) of Japan.

%

\end{document}